# Quantized Conductance in SnO$_2$ nanobelts with rectangular hard-walls

**Emilson R. Viana, Juan C. González, Geraldo M. Ribeiro and Alfredo G. de Oliveira.**

Quantized conductance is reported in high-crystalline tin oxide (SnO2) nanobelt back-gate field-effect transistors, at low temperatures. The quantized conductance was observed as current oscillations in the drain current vs. gate voltage characteristics, and were analyzed considering the nanobelt as a *quantum wire* with rectangular cross-section hard-walls. The quantum confinement in the nanowires created conditions for the successive filling of the electron energy-subbands, as the gate voltage increases. When the source-drain voltage is changed the oscillations are not dislocated with respect to V$_g$, indicating flat-band subband energies at low temperatures. The subband separation was found to be in good agreement with the experimental observations, since the oscillations tend to disappear for T > 60K. Therefore, a novel quantum effect is reported in SnO$_2$ nanobelts, which is expected to behave as bulk at zero electric gate fields.

*Universidade Federal de Minas Gerais, Instituto de Ciências Exatas, Departamento de Física, Belo Horizonte/MG, 31720-901, Brazil.*
*+55(31)34095625-34095637. E-mail: Corresponding author (E. R. Viana) emilsonfisica@ufmg.br*

## 1. Introduction

One dimensional nanostructures have stimulated significant contributions in scientific research in recent years, mostly due to interesting optical [1,2] and electrical [3,4] properties, related to their structural dimensionality and to quantum confinement effects [5]. Metal oxides semiconductor nanostructures, such as tin oxide (SnO$_2$) nanowires and nanobelts [6] have large potential for applications in gas [7,8] and ultraviolet light [9] sensors. In addition, functional oxides nanobelts can be used as an ideal system for a fully understanding of the space-confined transport phenomena[6].

Therefore, the study of the electrical transport in such nanostructures is important, in order to probe the low dimensionality and consequently the quantum effects in nano-based devices. However, the fundamental aspects of the electrical conduction mechanisms still remain unclear and challenging for SnO$_2$ nanobelts and nanowires. This lack of results is probably related to the hard task of metal-oxide nanowires characterization and to the difficulties in obtaining ohmic contacts [10,11].

Electronic transport phenomena in low-dimensional system can be divided basically into two categories: "ballistic transport", for very short wires, when the electrons can travel across the nanowires without any scaterring, and "diffusive transport", when the electrons suffer numerous scaterring events when travel along the wire, by phonos, boundary scaterring, lattice or other structured defects. For nanobelts sizes comparable with the electronic wavelength, the electronic density of states is altered drastically and quantum subbands are formed due to the quantum confinement effect at the boundary [26].

In this work, we show that high crystalline SnO$_2$ nanobelts at low temperatures when subjected to high electrical fields presents quantum confinement transport. The quantum confinement induces quantized conductance in the nanobelts, observed in the form of current oscillations in the drain current vs. gate voltage characteristics. The current oscillations were successfully explained by the successive filling of the electron energy-subbands, as the gate voltage increases, considering the nanobelt as a *quantum wire* with rectangular cross-section and hard-walls.

Despite of different approaches developed to study confinement effects [5,12,13,14], and ballistic transport in nanostructured materials [15,16], as far as we know, until the present date there is no report of the quantized conductance in the SnO$_2$ nanostructures that will be presented in the next sections.

## 2. Experimental Section

### 2.1 SnO$_2$ nanobelts synthesis

SnO$_2$ nanobelts were synthesized by using a gold-catalyst-assisted VLS method [17], based on thermal evaporation of oxide powders under controlled conditions. A suitable amount of pure Sn powder (1 g, 99.99% purity) was placed on top of a Si:SiO$_2$ substrate of 1x2 cm$^2$, previously coated with a 5 nm thick Au film that works as a catalyst. The substrate was placed inside of the horizontal quartz tube in a conventional tube furnace. The tube was firstly evacuated for 15 min, using a turbo-molecular vacuum pump to 10$^{-5}$ Torr, and then filled with pure argon. The furnace was then heated up to 800 ºC at a rate of 20 ºC/min, under a flux of argon. The temperature of 800 ºC was kept constant for 2-3 hours, and then the furnace was left to freely cool. At the end, SnO$_2$ nanobelts were found on the substrate surface in a cotton-wool-like form.

The morphology and crystal structure of the SnO$_2$ nanobelts were analyzed and characterized by scanning electron microscopy (SEM), transmission electron microscopy (TEM), atomic force microscopy (AFM), and X-ray diffraction (XRD). Figure 1 shows some morphological and structural investigations of the nanobelts. SEM examinations, see inset in fig.1.(a), reveal a large quantity of nanobelts with typical widths in the range of 50 to 500 nm, and lengths between 5 to 50 μm. TEM examinations, see

fig.1.(b), show that the nanobelts present a rectangular cross-section of about 25x180 nm$^2$, and are covered by a 10 – 20 nm thick amorphous tin oxide layer. Both, XRD and selected area electron diffraction (SAED), shown in fig.1.(a) and fig.1.(b), confirms the tetragonal rutile structure of the nanobelts with lattice constants of a = b = 0.473 nm and c = 0.318 nm [18].

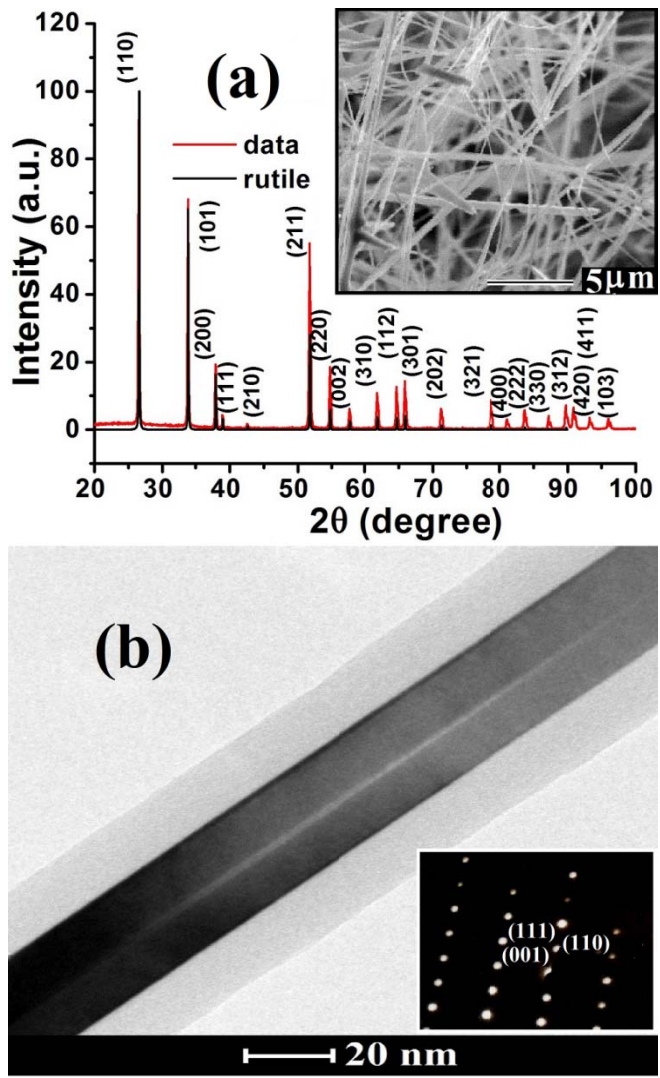

**Fig.1 (a)** XRD pattern of the nanobelts (red pattern) and the tetragonal rutile structure [6] (blue pattern). SEM image of several SnO$_2$ nanobelts is plotted in the inset. **(b)** TEM image of a SnO$_2$ nanobelt with rectangular cross-section covered by a thick amorphous layer. The SAED pattern of an individual nanowire is plotted in the inset.

**2.2 Nanobelt FET: device preparation**

In order to prepare nanobelt field effect transistor (NB-FET) devices, the SnO$_2$ nanobelts were dissolved in deionized (DI) water by sonication and then deposited on a SiO$_2$/Si substrate. NB-FET were fabricated by photolithography using a LaserWriter® (UV-Laser, λ=405 nm), and a standard lift-off process. Two-layer metallic electrodes, 2-3 μm apart and consisting of a 10/150 nm of Cr/Au bilayer, were defined on top of a 300 nm thick SiO$_2$ deposited on a high-doped p-type Si substrate.

Previously to the metal deposition, the high resistivity amorphous oxide layer surrounded the nanobelts, see fig.1.(b), was removed by a H$_2$-plasma treatment in order to improve the quality of the contacts. The plasma treatment on the surface of SnO$_2$ releases more charge carriers in the nanobelt conduction band by the creation of "oxygen vacancies" and less coordinated tin species [19,20]. We have found that at room temperature the electrical resistance of the non-treated SnO$_2$ nanobelts is higher than 10MΩ, and decreases to approximately 1MΩ after the plasma treatment. This enhance in the electrical conductivity of the nanobelts actually allows the low temperature measurements.

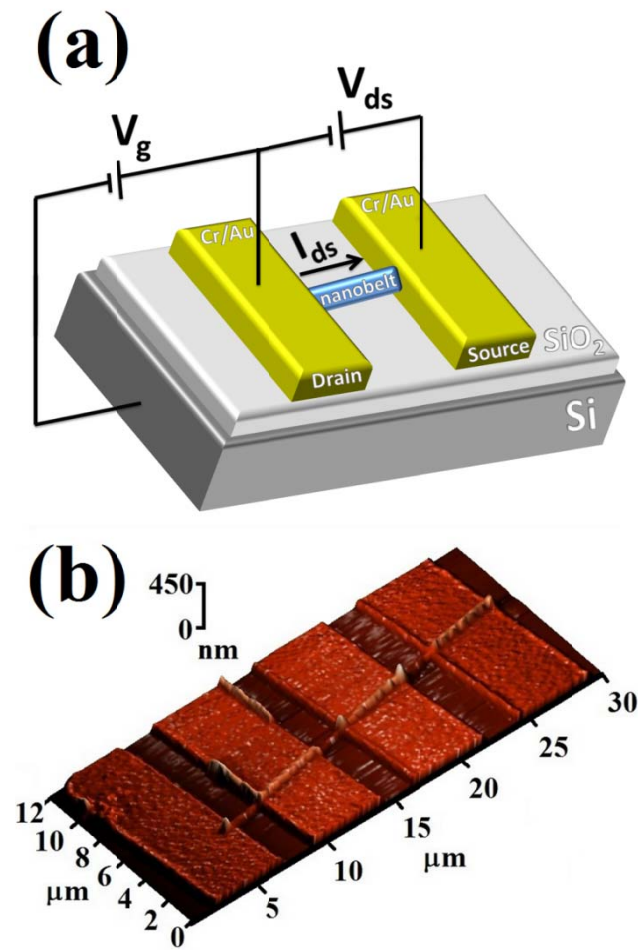

**Fig.2 (a)** Schematic back-gate FET-device used in this work. **(b)** AFM image of a NB-FET showing the 4 contacts on top of a SnO$_2$ nanobelt.

**3. Results/Discussion**

**3.1 Measurements**

The electrical transport study was carried out by connecting an individual nanobelt in a back-gate FET configuration and measuring their properties. The devices were placed in a cold-finger He-7 cryostat, model Oxford® CF1200, with a precise temperature-controller, model Oxford® ITC503. A Keithley® 237 source meter was used to applied and record the source-drain voltage ($V_{ds}$) and source-drain current ($I_{ds}$), respectively. The gate voltage ($V_g$) was applied by a Keithley® 230 voltage source.

The schematic of a NB-FET is shown in fig.2, together with an AFM image of the actual device. NB-FET device with four contacts were fabricated, but only the two central contacts were used in the measurements due to the high resistance of the nanobelts. The contact resistance was measured at low temperatures and represent less than 1% of the total resistance.

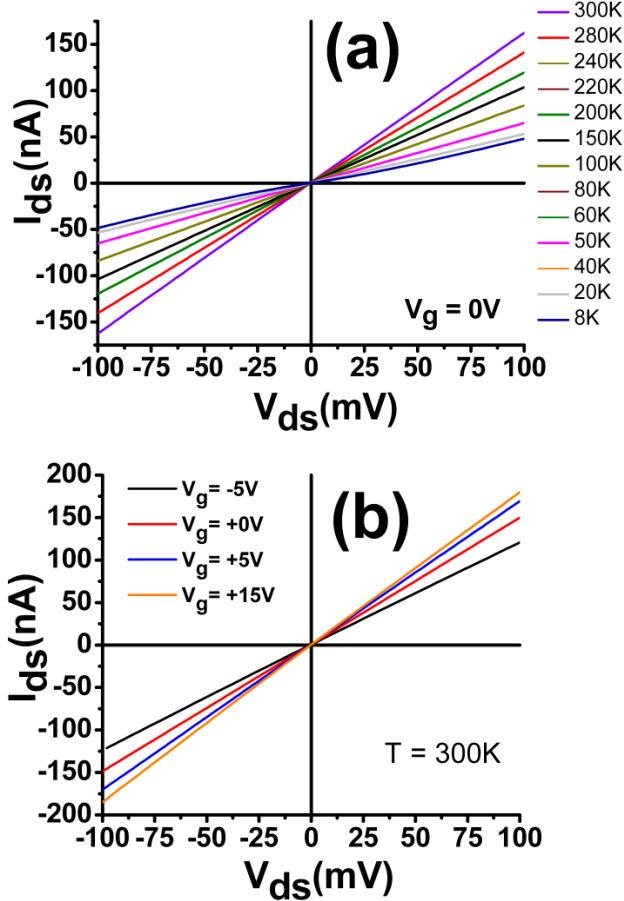

**Fig.3 (a)** $I_{ds}(V_{ds})$ curves as a function of temperature, with $V_g = 0$ V. **(b)** $I_{ds}(V_{ds})$ curves as a function of $V_g$ at T = 300 K.

In order to study the electrical transport properties of the $SnO_2$ nanobelts $I_{ds}$ was measured as a function of the source-drain voltage $I_{ds}(V_{ds})$, and the gate voltage $I_{ds}(V_g)$, for several temperatures T. Figure 3.(a) shows the $I_{ds}(V_{ds})$ curves as a function of T and $V_g$. For these measurements $V_{ds}$ was swap from -100 mV to +100 mV, in 1 mV steps. The linear behavior of the $I_{ds}$ curves shows good ohmic contacts, low contact resistance, and a transport dynamics governed by the $SnO_2$ nanobelt conduction channel. Besides that, the curves of fig.3.(a) show an increase of the conductance with temperature, as expect for a semiconductor material. The curves of fig.3.(b) show a reasonable field-effect-transistor behavior, that is, a transistor that relies on an electric field (gate-field) to control the shape, and hence the conductivity of a channel of charge carrier, in the semiconductor material.

Figure 4.(a) shows the $I_{ds}(V_g)$ curves as a function of $V_g$ in the temperature range from 6.6 to 50 K, with $V_{ds} = 100$ mV. This curves show ohmic behavior with a shift of the threshold voltage ($V_{th}$) towards more negative values, as the temperature increases. This behavior is the result of an increase in free carrier density with temperature [21].

At temperatures lower than 50 K some oscillations can be observed in the $I_{ds}(V_g)$ curves. This oscillations show that at low temperatures the electrical transport in the NB-FET exhibits quantized conductance. Several devices were investigated, showing more pronounced oscillations in thin nanobelts (25 nm < thickness < 35 nm, width ~ 200 nm) and disappear as the thickness, $L_y$, increases to approximately 45 nm.

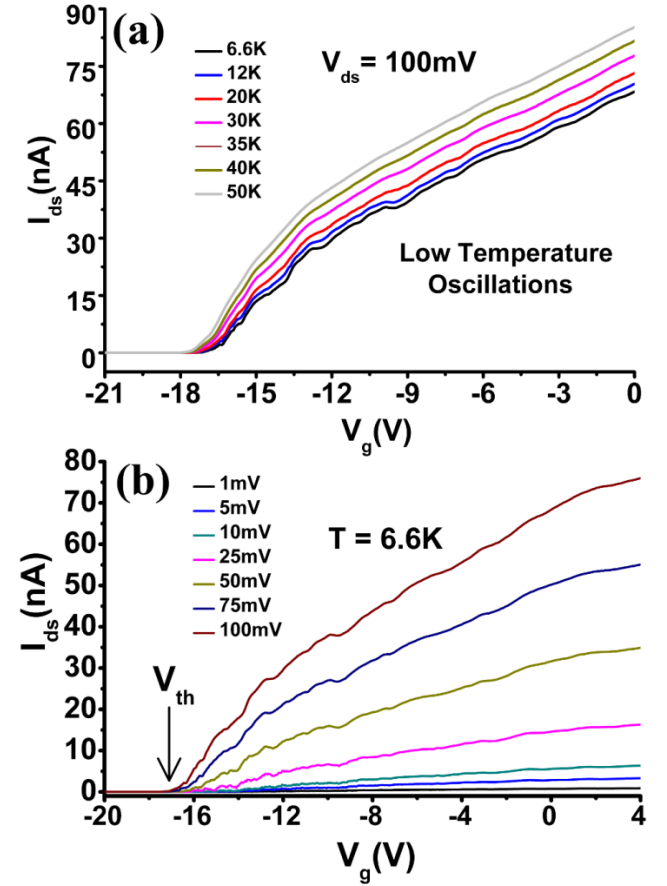

**Fig.4 (a)** $I_{ds}(V_g)$ curves as a function of temperature, in the range of 6.6 K to 50 K, and under constant bias voltage $V_{ds}$ =100 mV. At low temperatures, T < 50 K, a current staircase behavior is visible in the curves. **(b)** $I_{ds}(V_g)$ curves as a function of $V_{ds}$, at T = 6.6 K. Note the suppression of the current steps as $V_{ds}$ increases.

In figure 4.(b) the $I_{ds}(V_g)$ curves at T = 6.6 K as a function of $V_{ds}$ are presented. As $V_{ds}$ increases the current steps are suppressed, showing that the confinement effects gradually vanishes. The suppression of the current steps is a consequence of an increase of the electrons kinetic energy ($qV_{ds}$) above the energy spacing between the bottom of the subbands energy in the conduction band of the $SnO_2$ nanobelts.

The transconductance is the variation of the drain current $I_{ds}$ with gate voltage $V_g$, with a constant source-drain voltage $V_{ds}$ applied, and can be calculated by:

$$g_m = \left. \frac{d(I_{ds})}{d(V_g)} \right|_{V_{ds}=cte.} \quad (1)$$

In figure 5.(a) we show current oscillations in the $I_{ds}(V_g)$ curves, for $V_{ds}$ = 100mV and at temperature T(K) = 6.6K. The minimums in the $g_m$ curve, shown in fig. 5.(b), help us to identify more clearly the plateaus in $I_{ds}(V_g)$ of fig.5.(a). The plateaus or oscillations correspond to successive filling of the energy-subbands of the nanobelt. Twelve plateaus representing the energy-subbands where identified in $g_m$ curve, and the $V_g$ value of each plateaus was collected.

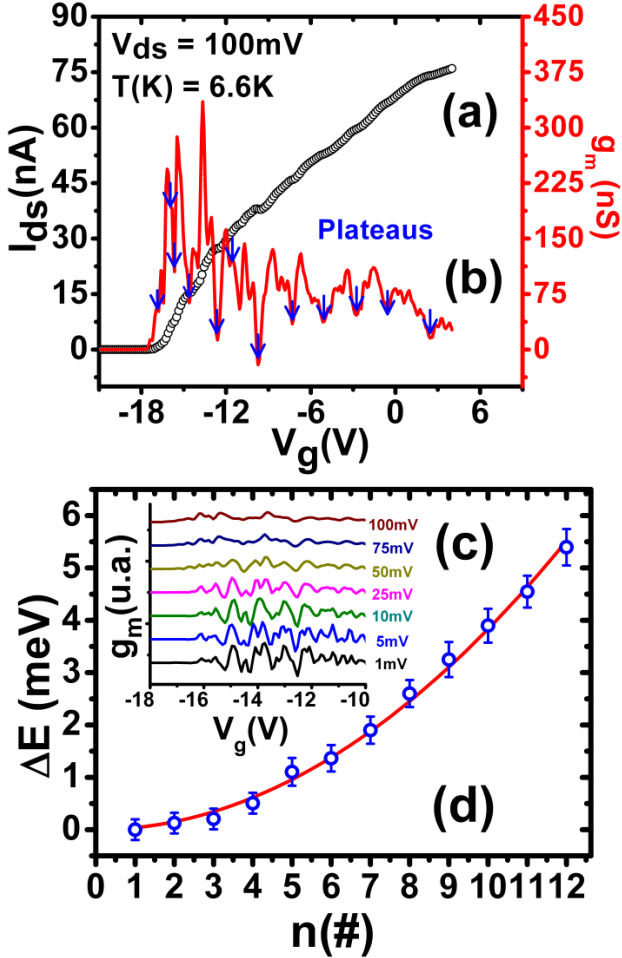

**Fig.5 (a)** $I_{ds}(V_g)$ curves for $V_{ds}$ = 100mV and at temperature T = 6.6K showing current oscillations. **(b)** The position $V_g$ of each peak identified in "(a)" was indexed to a energy level using eq.(3). **(c)** The transconductance $g_m$ calculated using eq.(1), for measurements at T = 10K, as a function of $V_{ds}$. **(d)** Energy levels plotted against the quantum number $n$ showing quadratic dispersion, the results are fitted using eq.(2).

### 3.2 Theoretical and Experimental Analysis

In a NB-FET with a small cross-section, conduction occurs in the innermost layer of the nanobelt. For values of $V_g$ below $V_{th}$ no conduction takes place from source to drain because of the "electrostatic squeezing" [16] of the conduction channel. For gate voltages just above the threshold, conduction is measured and the carriers are confined within a filament conduction channel from source to drain. The filament cross-section increases with the gate voltage, until the flat band behavior is reached, that is when the filament cross-section becomes equal to the physical cross-section of the nanobelt [5].

By solving an ideal 2D rectangular hard-potential, the electronic structure of the nanobelt can be calculated. In this *quantum wire model*, the electron energy levels, with respect to the bottom of the bulk conduction band, can be calculated by [24]:

$$E_{yz}(n_y, n_z) = \frac{\pi^2 \hbar^2}{2m^*_{y,z}} \left( \frac{n_y^2}{L_y^2} + \frac{n_z^2}{L_z^2} \right) \quad (2)$$

where $E_{yz}$ is the edge of the subbands. The quantum numbers $n_y$ and $n_z$, defined by non-zero integers, stand for the indexes of the energy subbands, and $L_y$ and $L_z$, represent the thickness and width of the nanobelt, respectively. The effective mass of the electrons in $SnO_2$ is $m^*_{y,z} = 0.3 m_o$ [19]

The experimental energy separation, $\Delta E_{exp}$, between the two subbands can also be estimated from the values of the gate voltage separation $\Delta V_g$ between two adjacent oscillations [12,25] by:

$$\Delta E_{exp} = \frac{(C/A).\Delta V_g.h^2}{8\pi.e.m^*} \quad (3)$$

The capacitance per unit area can be estimated by the parallel-plane capacitor model, $(C/A) = (\varepsilon_{SiO_2}/d) = 115\mu F/m^2$, were $(\varepsilon_r.\varepsilon_o) = \varepsilon_{SiO_2}$ is the dielectric constant substrate oxide layer $SiO_2$ ($\varepsilon_r = 3.9$), and $d$ = 300nm is the thickness of the $SiO_2$ layer. Equation (3) comes from the fact that the electron concentration obtained by the integration of the electron density of states, should be equal to that induced by the applied gate voltage [12].

From the $V_g$ values obtained in the transconductance curve $g_m$ (fig.5.(b)), and using eq.(3), the experimental subband energy separation between the subbands identified $\Delta E_{exp}$ were estimated. Since the theoretical model used to explain the *quantum wire* phenomena is eq.(2), we have plotted the energy levels $\Delta E_{exp}$ with respect to the occupation number $n$ as presented in fig.5.(d).

Figure 5.(d) shows the energy separation between the twelve observed plateaus with respect to the plateau at the lowest $V_g$. Clearly the quadratic dependence of $\Delta E_{exp} = a.n^2$, with respect to their index number $n$ was found, suggesting that the plateaus are caused by the consecutive filling of the energy-subbands in the nanobelt. The energy separation between the first and last plateaus is of less than 6meV, indicating that only the energy-subbands with $n_y$ =1 and $n_z$ = 1 to 12 were filled for the nanobelt with $L_y$ = 25nm. From eq.(2) the coefficient $a$ of the fitting of $\Delta E_{exp}$ curve with $(a.n^2)$ is $\left( \pi^2 \hbar^2 / 2 m^*_{y,z} L_z^2 \right)$, given a value of $L_z$ = 180nm in very good agreement with the width of the nanobelt, measured by AFM.

Our simple theoretical *quantum wire* model well explains the energy separation between consecutives plateaus or oscillations in the $I_{ds}(V_g)$ curve of the nanobelt. Therefore the oscillations in the $I_{ds}(V_g)$ corresponds to the consecutive filling of energy-subbands as the applied gate voltage rises up the Fermi level though the electronic structure of the nanobelt. Furthermore the observation of current oscillations only for temperatures below 60K (5.2meV) are consistent with the fact that the quantized conductance is only visible in the $I_{ds}(V_g)$ curve if the thermal energy ($k_B T$) is lower or comparable with the highest subband energy that is of the order of $[E_{yz}(2,1) - E_{yz}(1,1)] = 6meV$ for our nanobelt. When the source-drain-voltage $V_{ds}$ was changed in the $I_{ds}(V_g)$ curves at 6.6K, the oscillations, identified as plateaus in the transcondutance curves of fig.5.(c), were not dislocated in respect with $V_g$ axis indicating flat-band subband energies.

## 4. Conclusions

In conclusion, the electrical transport mechanism of $SnO_2$ nanobelts was investigated by building and characterizing single nanobelts in a back-gate field-effect transistor (NB-FET) configuration. The NB-FET shows quantized conductance due to quantum-confined transport phenomena, that was observed in the form of oscillations in the drain-source current *vs.* gate voltage characteristics, $I_{ds}(V_g)$. The observed current oscillations were explained by the successive filling of electron energy subbands as the gate voltage increases.

The transconductance $g_m$ was also calculated from the $I_{ds}(V_g)$ curve and show several minimums that helped us to identify the plateaus in the $I_{ds}(V_g)$ curve and consequently the effect of energy-subband separation, more clearly. With that, we estimated the experimental subband energy separation $\Delta E_{exp}$ between consecutive subbands and is clearly show a quadratic dependence of $\Delta E_{exp}$ with the index number $n$.

The theoretical difference between the subband energy-levels with respect to the ground-state is in agreement with the behavior of the nanobelt as a *quantum wire* with rectangular cross-section and hard-walls. All detected subbands have quantum numbers $n_y$ = 1, and $n_z$ < 12 for the range of temperature and applied voltages, here studied. We could estimate the maximum subband energy separation of 6 meV that is also in agreement with the detected threshold-temperature of the current oscillations, $T_{th}$ = 60K, experimentally determined.

Within the analysis present and discussed here, it was estimated in which circumstances the phenomena of quantized conductance can be observed in the $SnO_2$ nanobelts, enabling the experimental evaluation of the subband energy separation of the nanobelt that refers to a "subband spectroscopy" of the electronic structure [5].

Despite of several different approaches developed to study confinement effects [5, 12, 13, 14] and ballistic transport in nano-structured materials [15,16], as far as we know, until the present date there is no report of the quantized conductance in $SnO_2$ nanobelts that was presented and analyzed in details here.

## Acknowledgments

The authors' thanks Prof. Dr. Bernardo Ruegger Almeida Neves, and Dr. Ana Paula Barbosa for the presented AFM images. Prof. Dr. Nivaldo Lúcio Speziali for the XRD measurements, Prof. Dr. Além-Mar Bernardes Gonçalves for invaluable discussion on transport measurements and device fabrication, and the technical support of the Microscopy Center of the Universidade Federal de Minas Gerais. Thanks CNPq, CAPES and FAPEMIG, Brazilian official agencies for funding this work.